\documentclass{ipi}
\usepackage{amssymb}
\usepackage{amsmath}
\usepackage{multirow}
\usepackage[noend]{algpseudocode}
\usepackage{amsfonts}       
\usepackage{nicefrac}       
\usepackage{microtype} 
\usepackage[utf8]{inputenc} 
\usepackage[T1]{fontenc}    

\usepackage{gensymb}
\usepackage{textcomp}
\usepackage{xcolor}

\usepackage{subfigure}
\usepackage{dsfont}
\usepackage{multirow} 
\usepackage[pagewise]{lineno}
\usepackage{graphicx}
 \usepackage[colorlinks=true]{hyperref}
\hypersetup{urlcolor=blue, citecolor=red}

\def\currentvolume{X}
 \def\currentissue{X}
  \def\currentyear{200X}
   \def\currentmonth{XX}
    \def\ppages{X--XX}
     \def\DOI{10.3934/xx.xxxxxxx}

\theoremstyle{definition}

\title
      [Learning to Scan for CT Imaging]{Learning to Scan: A Deep Reinforcement Learning Approach for Personalized Scanning in CT Imaging}

\author[Ziju Shen and Yufei Wang and Dufan Wu and Xu Yang and Bin Dong]{}


 \email{zjshen@pku.edu.cn}
 \email{yufeiw2@andrew.cmu.edu}
 \email{dwu6@mgh.harvard.edu}
 \email{xuyang@math.ucsb.edu}
 \email{dongbin@math.pku.edu.cn}

\thanks{* Equal Contribution}
\thanks{$\dagger$ Corresponding Author}
\begin{document}
\maketitle
\centerline{Ziju Shen$^*$}
{\footnotesize
 \centerline{Academy for Advanced Interdisciplinary Studies}
 \centerline{Peking University, P.R. China}
} 

\medskip

\centerline{Yufei Wang$^*$}
{\footnotesize
 \centerline{Computer Science Department}
  \centerline{Carnegie Mellon University, USA}
}
\medskip
\centerline{Dufan Wu}
{\footnotesize
 \centerline{Center for Advanced Medical Computing and Analysis}
  \centerline{Massachusetts General Hospital and Harvard Medical School, USA}
}

\medskip
\centerline{Xu Yang}
{\footnotesize
 \centerline{Department of Mathematics}
  \centerline{University of California, Santa Barbara, USA}
}
\medskip
\centerline{Bin Dong$^\dagger$}
{\footnotesize
 \centerline{Beijing International Center for Mathematical Researchs}
  \centerline{Center for Data Science, Institute for Artificial Intelligence}
  \centerline{Peking University, P.R. China}
}

\begin{abstract}
Computed Tomography (CT) takes X-ray measurements on the subjects to reconstruct tomographic images. As X-ray is radioactive, it is desirable to control the total amount of dose of X-ray for safety concerns. Therefore, we can only select a limited number of measurement angles and assign each of them limited amount of dose. Traditional methods such as compressed sensing usually randomly select the angles and equally distribute the allowed dose on them. In most CT reconstruction models, the emphasize is on designing effective image representations, while much less emphasize is on improving the scanning strategy. The simple scanning strategy of random angle selection and equal dose distribution performs well in general, but they may not be ideal for each individual subject. It is more desirable to design a personalized scanning strategy for each subject to obtain better reconstruction result. In this paper, we propose to use Reinforcement Learning (RL) to learn a personalized scanning policy to select the angles and the dose at each chosen angle for each individual subject. We first formulate the CT scanning process as an Markov Decision Process (MDP), and then use modern deep RL methods to solve it. The learned personalized scanning strategy not only leads to better reconstruction results, but also shows strong generalization to be combined with different reconstruction algorithms.

\end{abstract}

\section{Introduction}
\label{sec:intro}
X-ray based computed tomography (CT) is a medical imaging procedure that reconstructs tomographic images by taking 
X-ray measurements from different angles. 
To obtain high-quality reconstructions, in early reconstruction algorithms such as filtered backproject (FBP)~\cite{FBP} and algebra reconstruction technique (ART)~\cite{ART}, a number of different angles need to be measured. However, 
since X-ray is radioactive, the total dose of X-ray needs to be restricted in the scanning  process, and thus we need to either decrease the X-ray intensity at each chosen angle, or to reduce the total number of angles taken.  Decreasing X-ray intensity in each angle will result in more noisy measurements, while fewer angles will reduce the information we need for a high-quality reconstruction. This causes great challenges in designing efficient and effective reconstruction algorithms.
  
Compressed sensing~\cite{compress_sense} resolves the issue to a certain extend. According to the theory of compressed sensing, if an image has a sparse property after certain transformations (e.g., wavelet transform), then it can be robustly reconstructed with a reduced number of random measurements by solving an $l_1$-minimization problem when the measurements and the transformation satisfy the D-RIP condition~\cite{drip}. We can use the alternating direction method of multipliers (ADMM) \cite{admm1,admm2,cai2009split} or primal-dual hybrid gradient method (PDHG) \cite{pdhg1,pdhg2,pdhg3} to solve this $l_1$-minimization problem to obtain a reconstructed image.

In the literature of CT image reconstruction or image restoration in general, people focus on designing effective regularizations, which includes the total variation (TV)~\cite{rudin1992nonlinear}, nonlocal means~\cite{buades2005non},
block-matching and 3-Dfiltering (BM3D)~\cite{dabov2007image},  weighted nuclear norm minimization (WNNM)~\cite{gu2014weighted}, wavelets and wavelet frame models~\cite{daubechies1992ten, mallat1999wavelet, dong2010mra}, K-SVD~\cite{elad2006image}, data-driven (tight) frame~\cite{cai2014data, tai2016multiscale}, 
low dimensional manifold method (LDMM)~\cite{osher2017low}, etc. More recently, the rapid development of machine learning,
especially deep learning, has lead to a paradigm shift
of modeling and algorithmic design in computer vision
and medical imaging~\cite{wang2016perspective, wang2017machine, mccann2017convolutional, wang2018image, zhang2020review}. Deep learning based models are able to leverage large image datasets to learn better image representations and produce better image reconstruction results than traditional methods~\cite{chen2017low, jin2017deep, kang2017deep, zhang2018sparse, yang2018low, shen2018intelligent}.

In most CT reconstruction models, the emphasize is on designing effective image representation, while much less emphasize is on improving the scanning strategy.
In compressed sensing, the scanning strategy is entirely random~\cite{rip, drip}, i.e., the measurement angles are selected randomly and the dose are allocated uniformly across the angles. In theory, such random sampling is proven for exact recovery using a convex model for MRI. However, such result is generally untrue for CT imaging due to the coherence structure of Radon transform. In practice, uniform sampling is often adopted. However, for each individual subject, a uniform or random scanning strategy may not be ideal. It is more desirable to design a personalized scanning strategy for each subject to achieve better reconstruction results. Our key observation is that the measurements collected in the early stage  during the scanning process can be used to guide the later scanning. 



Despite the potential improvement of a personalized 
scanning strategy for each individual subject, it is very difficult to handcraft such a strategy by a human expert.  This is where machine learning can help. The personalized scanning strategy can be learned using either active learning~\cite{settles2009active} or Reinforcement Learning (RL)~\cite{RLoverview}. 
In this paper, we propose to use reinforcement learning to learn such a personalized scanning strategy for each subject. The reason we choose RL over active learning is that RL is non-greedy and naturally guarantees the long-term reconstruction quality. We formulate the CT scanning process as a Markov Decision Process (MDP), where the state includes currently collected measurements, the action determines the next measurement angle and the dose usage, and the reward depends on the reconstruction quality. We further use modern Deep RL algorithms to solve it.  We show in the experiments that the personalized scanning policy learned by RL significantly outperforms the random scanning strategy in terms of the reconstruction quality, and can generalize to be combined with different reconstruction algorithms. 

We note that current commercial CT system does not support to select angles and dose as freely as what the trained RL agent suggests since it requires fast and irregular acquisition motion of the gantry. The proposed approach serves as a proof of concept and experimental study to investigate the potential benefits of the scanning strategies suggested by the trained RL agents. Nonetheless, there are multi-beam X-ray systems available to which the proposed scanning strategy can be applied to \cite{zhang2006multi,multisource,park2020multi}.
  
  \subsection{Related works}
  

  For compressed sensing, there have been two primary categories of scanning strategies: static and dynamic. Static scanning strategy refers to the method which collects measurements in a fixed order. Low-discrepancy sampling~\cite{low_discrepancy} and uniformly spaced sparse sampling methods~\cite{uniform_sample} are two examples of static scanning strategy. Non-uniform static scanning strategy based on the model of the subject to be scanned is proposed in~\cite{model_sample,model_sample2}. However, because the order of measurements is predetermined, static scanning strategy is not flexible for different subjects and may 
  lead to poor results for some of them.
  
  Dynamic scanning strategy refers to the methods which collect measurement adaptively based on information obtained from previous measurements. One traditional method tries to find the most suitable measurements which can minimize the entropy to decrease uncertainty of images, such as BCS~\cite{BCS,CSBayesian}. Similarly, other methods \cite{inform_gain,DS_information_gain} use the information gained at each additional scan to guide the selection of the next measurement. However, these methods are typically greedy methods in nature, have many hyperparameters to be properly tuned, and are slow during inference as they either need to take inverse of large matrices, or to run the reconstruction algorithm for many times when determining the best next angle.
  More recently, deep neural networks are used to estimate the expected reduction in distortion (ERD) in the reconstructed image when an additional measurement is selected~\cite{SLAD2,SLADS,SLADS2,SLADS3,SLADS4,SLADS5}. However, for the estimate of ERD to be accurate, it requires a large number of measurements in training. 
  
  All the above methods are not specific to CT scanning. They are greedy in nature and do not provide a strategy for dose allocation. In contrast, RL is able to generate a non-greedy policy that aims at maximizing long-term rewards which, in this paper, is the quality of the reconstructed CT image. Furthermore, the setting of RL is flexible enough to handle both angle selection and dose allocation, and even more decision options during scanning. Therefore, in this paper 
  we use RL to design a scanning policy that acts optimally on each individual subject. In Scanning Transmission Electron Microscopy (STEM), a recent work by~\cite{RLSTEM} proposes to use RL to guide the movement of the detector and uses a generator to generate reconstructed images. However, since the image modality is drastically different from CT, the proposed MDP (especially the state, action and architecture of the policy network) is vastly different from what is proposed in this paper. 
  
  Recently, there has been a line of work that uses RL to solve combinatorial optimization problems, such as Travelling Salesman Problem~\cite{bello2016neural}, Vehicle Routing Problem~\cite{kool2018attention}, Influence Maximization~\cite{mittal2019learning}, Autonomous Exploration~\cite{auto_explore}, and they all show RL can obtain better results than traditional solvers. Given the success of these prior works, and that the angle selection in CT scanning is also a combinatorial optimization problem, it naturally motivates us to try to use RL to solve the problem and see if it brings any benefit.
  

\section{Preliminaries}
\label{sec::preliminaries}

\subsection{A Brief Review on MDP and Reinforcement Learning~\cite{sutton2018reinforcement}}
A sequential decision problem can be formulated as a Markov Decision Process (MDP).
MDP is a tuple $(\mathcal{S},\mathcal{A},\gamma,\mathbb{P},r)$ that consists of the state space $\mathcal{S}$, the action space $\mathcal{A}$, the discount factor $\gamma$, the transition probability of the environment $\mathbb{P}: \mathcal{S} \times \mathcal{A} \times \mathcal{S} \rightarrow [0, 1]$ and the reward $r:  \mathcal{S} \times \mathcal{A} \rightarrow \mathcal{R}$. 
A policy $\pi$ in RL is a probability distribution on the action $\mathcal{A}$ over $\mathcal{S}$: $\pi: \mathcal{S} \times \mathcal{A} \rightarrow [0, 1]$.
Denote the interactions between the agent and the environment as a trajectory $\tau = (s_0, a_0, r_0, ..., s_T, a_T, r_T, ...)$. The return of $\tau$ is the discounted sum of all its future rewards:
$$G(\tau) = \sum_{t=0}^{\infty}\gamma^{t}r_{t}.$$
Given a policy $\pi$, the value of a state $s$ is defined as the expected return of all the trajectories when the agent starts at $s$ and then follows $\pi$: 
$$V^{\pi}(s) = E_{\tau}[G(\tau) | \tau(s_0) = s, \tau \sim \pi]$$ 
Similarly, the value of a state-action pair is defined as the expected return of all trajectories when the agent starts at $s$, takes action $a$, and then follows $\pi$: $$Q^{\pi}(s,a) = E_{\tau}[G(\tau) | \tau(s_0) = s, \tau(a_0) = a, \tau \sim \pi]$$

Given an MDP, the goal of a reinforcement learning algorithm is to find a policy $\pi$ that maximizes the discounted accumulated rewards in this MDP: 
\begin{equation}
\begin{split}
\max_{\pi}\eta(\pi)=\mathbb{E}_{s_0 \sim \rho(s))}[V^\pi(s)].
\end{split}
\end{equation}
Many effective RL algorithms have been developed to find the optimal policy $\pi$. 
They can be generally classified into two categories: valued-based methods and policy gradient methods. 

Value-based methods such as Q-learning~\cite{watkins1992q, DQN} uses the Bellman Equation to learn the optimal Q function, and then derive the optimal policy by acting greedily according to the optimal Q function. Formally, the Bellman Optimal Equation is:
\begin{equation*}
    \begin{split}
        Q^*(s, a) &= \max_\pi Q^\pi(s,a) \\
        Q^*(s_t, a_t) &= r(s_t, a_t) + \gamma \mathbb{E}_{s_{t+1} \sim P(\cdot | s_t, a_t)} \left[\max_{a_{t+1}} Q^*(s_{t+1}, a_{t+1}) \right]
    \end{split}
\end{equation*}

Methods like DQN~\cite{mnih2015human} uses a deep neural network $Q_w$
to
learn the optimal $Q^*$ by
storing transitions $\{s, a, r, s'\}$ in an off-line replay buffer
and minimizing the following Bellman Error:
\begin{equation*}
L(w) = \mathbb{E}_{s,a,r,s'} [(Q_w(s, a) - (r + \gamma \max_{a'}
Q_{\tilde{w}}(s', a'))^2] 
\end{equation*}
where $Q_{\tilde{w}}$
is the target Q network whose parameters $\tilde{w}$ are
periodically copied from $w$ to stabilize training. After the optimal Q function is learned, the policy is simply $\pi(s) = \arg\max_a Q_w(s, a)$. Due to this $\arg\max$ operator, value-based methods are mostly suitable for discrete actions. Because our action contains continuous variables (the dose allocation), we do not use value-based methods as our RL algorithm.

Policy gradient methods~\cite{sutton2000policy, silver2014deterministic} directly optimize a parameterized policy by computing a surrogate objective. Given a policy $\pi_\theta$ parameterized by a neural network $\theta$, the policy gradient theorem~\cite{sutton2000policy} states that:
\begin{equation*}
\nabla_\theta \eta(\pi_\theta) = \mathbb{E}_{s,a\sim\pi_\theta}
[\nabla_\theta \log\pi_\theta(a|s) \cdot Q^{\pi_\theta}
(s, a)]
\end{equation*}
Policy gradient methods is suitable for continuous action space. One drawback of vanilla policy gradient method is that the gradient might have high variance and make the update of $\theta$ unstable. Many more advanced methods have then been proposed to address this.  In this paper, we use the Proximal Policy Optimization (PPO)  algorithm ~\cite{ppo}, which updates the policy in a proximal region to avoid unstable updates.
We now briefly review how it works.
Given a parameterized policy $\pi_{\theta}$,  its advantage function is defined as 
$A^{\pi_\theta}(s,a)=Q^{\pi_\theta}(s,a) - V^{\pi_\theta}(s)$.
Given an old policy $\pi_{\theta_{old}}$, let $b_\theta(s, a) = \frac{\pi_\theta(a|s)}{\pi_{\theta_{old}}(a|s)}$, PPO optimizes $\pi_{\theta}$ w.r.t. the following surrogate objective using gradient descent:
\begin{equation}
    J^{\text{PPO}}(\theta)=\mathbb{E}_{s,a\sim\pi_{\theta_\text{old}}}\left[\min(b_\theta(s, a) A^{\pi_{\theta_{old}}}(s, a), \text{clip}(b_\theta(s,a), 1-\varepsilon,1 + \varepsilon)A^{\pi_{\theta_{old}}}(s, a))\right]
    \label{eqn:ppo_J},
\end{equation}
where $\text{clip}(x,a,b)=\max(\min(x,b),a)$.

\subsection{CT Reconstruction}
One of the common CT systems is the cone-beam CT system. In the 2-dimensional case, it is known as the fan-beam CT, and this is the CT system that we focus on in this paper. For a given angle $\theta$ and X-ray beamlet $r$, the X-ray projection operator $A^{\theta,r}$ is defined as follows:
\begin{equation}
A^{\theta,r}[f](t)=\int_{0}^{L(t)}f(\mathbf{x}_{\theta}+\mathbf{n}l)dl
\end{equation}
where $f$ is the unknown image (X-ray attenuation coefficients) that needs to be reconstructed, $\mathbf{x}_\theta=(x_\theta,y_\theta)$ represents the coordinate of the X-ray source which is different for different projection angle $\theta$, $\mathbf{n}=(n_x,n_y)$ is the direction vector of beamlet $r$, $t$ is the coordinate on the X-ray imager which is precisely the intersection of the beamlet $r$ with the X-ray imager. $L(t)$ is the length of the X-ray beamlet from the source to the location $t$ on the imager. If $A^{\theta,r}[f](t)$ is sampled with respect to $t$ for each angle $\theta$, the resulting data projection can essentially be written as a vector $p_{\theta}$. Now, putting the vectors $p_\theta$ together for all different angles $\theta$, we obtain an image denoted as $p$ whose columns are formed by $p_\theta$. 

We can write the CT image reconstruction problem as a linear inverse problem
\begin{equation}
    p=Af+\epsilon,
    \label{equ:CTmeasurement}
\end{equation}
where $A$ is the linear operator represents the collection of discrete line integration at different projection angles and along different beamlets,  and
$\epsilon$ is an additive noise.
In our simulations, the matrix $A$ is generated by Siddon’s algorithm~\cite{FastCT1985} 
As equation \eqref{equ:CTmeasurement} is a linear equation, it can be directly solved by ART or SART algorithm~\cite{ART, SART}. However, as $m$ can be far smaller than $n$, equation \eqref{equ:CTmeasurement} has far less equations than unknowns. 
In order to obtain high-quality solutions, regularization-based models are often used, which typically take the form as follows:
\begin{equation}
        \min_{f}\frac{1}{2}\| p-Af\|_2^2+\lambda R(f),
\end{equation}

where $R(f)$ is the regularization term. Two benchmarking regularization terms are TV regularization $R(f)=\|\nabla f\|_1$~\cite{TVmodel1} and wavelet regularization $R(f)=\|Wf\|_1$~\cite{WaveletFrame}, where $W$ is the wavelet transform. 
Both of these two optimization problems can be solved by ADMM or PDHG. 


\subsection{Relationship between Measurement Noise and Dose}
\label{sec:noise}
Noise intensity on measurements heavily relies on the X-ray dose. It is common to assume that the measurement noise follows a Gaussian distribution\cite{CT_sim},  $\epsilon\sim\mathcal{N}(0,\sigma)$, and
\begin{equation}
    \sigma\varpropto\frac{1}{\sqrt{n_{\max}d\exp{(-P)}}},
\end{equation}
where $d$ is the X-ray dose used in a measurement, $P$ is the average intensity of measurement, and $n_{max}$ is the maximum number of photons the source can generate. We can easily see that if we use more dose, the noise level becomes smaller. We note that the Poisson distribution can also be used here. However, according to~\cite{CT_sim}, the Gaussian distribution approximates the noise well enough.
 
 \subsection{Some Further Discussions}
 As equation \eqref{equ:CTmeasurement} shows, the measurements we obtain from a CT scan depends both on the angle (which determines $A$), and the X-ray dose (which determines $\epsilon$). Due to the limitation on X-ray dose usage, we can only select a limited number of angles and assign each of them limited amount of dose. Traditional methods simply randomly select the angles and equally distribute the allowed dose on them. Our goal is to use RL to learn a personalized policy to select the angles and the dose at each chosen angle for each individual subject.
 

\section{Method}
\label{sec::algo}
Our goal is to learn a policy that can decide the next measurement angle and its corresponding X-ray dose based on the measurements that we have already obtained in the scanning process. We now present how the scanning process can be formulated as a Markov Decision Process (MDP) and solved by reinforcement learning algorithms. 

\subsection{MDP Formulation of Personalized Scanning}
We note that the angle selection problem in CT scanning itself is a combinatorial optimization problem and is NP-Hard. However, we can view it as a sequential decision process and use RL to solve it effectively. This is similar to works that use RL to solve other combinatorial optimization problems such as Influence Maximization~\cite{mittal2019learning}. Specifically,  
the CT scanning process can be viewed as a sequential decision process, where at each time step we need to decide on the measurement angle and the corresponding X-ray dose. 
Given an Image $I$ and the number of all possible angles $N$ (e.g., $N = 360$ if we can choose all the integer angles from $0$\textdegree to $359$\textdegree), we now elaborate how the CT scanning process on $I$ can be formulated as an MDP:
\begin{itemize}
    \item[1)] The \textbf{state} is a sequence
    $\vec{s_t}=(s_1,s_2,...,s_t)$, where $s_t=(p_t, d_t^{\text{ac}},d_t^{\text{rest}})$. $p_t$ is the collected measurement at time step $t$. $d_t^{\text{ac}}$ records the used dose distribution up to time step $t$. It is an $N$ dimensional vector, and the value at each entry represents the used X-ray dose at that corresponding angle. $d_t^{\text{rest}}$ is a scalar that
    represents the amount of the remaining dose that we can use. Because the reconstructed image at time step $t$ relies on all previously collected measurements, we include all of them in the state.
    
    \item[2)] The \textbf{action} is $a_t=(a_{t}^\text{angle},a_{t}^{\text{dose}})$. $a_{t}^\text{angle}$ is the angle we choose at time step $t$, and it is a one-hot vector of dimension $N$. $a_{t}^{\text{dose}} \in [0,1]$ is the fraction of dose that we apply at the corresponding angle. If at a certain time $t$, the total used dose exceeds the total allowed dose, we clip the exceeding dose and terminate the MDP. 
    \item[3)] The \textbf{reward} is computed as  $r(s_t, a_t) = \text{PSNR}(I_t, I)-\text{PSNR}(I_{t-1}, I)$, where $I$ is the groundtruth image, $I_t$ is the reconstructed image at time step $t$, and $\text{PSNR}(\hat{I}, I)$ represents the Peak Signal to Noise Ratio (PSNR) value of the reconstructed image $\hat{I}$. We use the increment of PSNR to evaluate how much benefit the new chosen angle/dose brings. The reconstructed image $I_t$ can be obtained from any reconstruction algorithm such as
    SART, TV regularization \cite{rudin1992nonlinear,goldstein2009split}, wavelet frame regularization \cite{ron1997affine,cai2009split}, and any modern deep learning based methods~\cite{pdhgnet}.
    Note that a more refined image reconstruction algorithm leads to higher quality reconstructed image which may bring benefit for training the RL policy. However, regularization based methods normally have at least one hyerparameter that needs to be tuned when number of projections changes for optimal reconstruction. On the other hand, deep learning methods normally have generalization issue when number of projections vary drastically. Therefore, for  practical concern, it is more convenient to apply SART or other iterative reconstruction methods that do not have generalization issues and do not require hyperprameter tuning. Furthermore, angle selection and dose allocation may not require very refined reconstructions, but rather a global reconstruction of anatomical structures. Therefore, we shall use SART to compute the reward.
    \item[4)] The \textbf{transition model} $\mathbb{P}$ represents the scanning process of CT. 
    At time step $t$, given the state $\vec{s}_t$ and action $a_t$, the next state $\vec{s}_{t+1}$ is simply the concatenation of $\vec{s}_t$ and $s_{t+1} = (p_{t+1}, d_{t+1}^{\text{ac}}, d_{t+1}^{\text{rest}})$. We now show how each of the three elements in $s_{t+1}$ can be computed. 
    \begin{enumerate}
        \item The new measurement is obtained as $p_{t+1} = p_{t+1}^{\text{true}} + \epsilon$, where $p_{t+1}^{\text{true}}$ is the clean projected value obtained using the chosen angle $a_t^{\text{angle}}$, and $\epsilon$ is the measurement noise. The noise depends on the chosen dose $a_t^{\text{dose}}$ as mentioned in section \ref{sec:noise}: $\epsilon\sim \mathcal{N}(0,\sigma)$,  $\sigma\varpropto\frac{1}{\sqrt{n_{\max} a_t^{\text{dose}}\exp{(-P)}}}$, where $P$ is the average of $p_{t+1}^{\text{true}}$.
        \item The new dose distribution is obtained by adding the new decision: $d_{t+1}^{\text{ac}} = d_{t}^{\text{ac}} + \mathds{1}_{a_t^{\text{angle}}} \cdot a^{\text{dose}}_t$, where $\mathds{1}_{a_t^{\text{angle}}}$ is the one-hot vector of the chosen angle $a_t^{\text{angle}}$.
        \item The rest amount of dose is updated by subtracting the used dose: $d_{t+1}^{\text{rest}} = d_{t}^{\text{rest}} - a_t^{\text{dose}}$. The MDP terminates once the dose is used up.
    \end{enumerate} 
\end{itemize}
As we choose the increment in PSNR as the reward, the total sum of reward (when there is no discounting, as in our experiments) is the PSNR value of the  final reconstructed image. Therefore, if we find the optimal policy to this MDP, it will also have the best reconstruction result for the image.

We note that, the final output of the RL policy is a set of angles (as well as the corresponding allocated doses). Although the angles are selected by the RL policy in an orderly fashion, in terms of image reconstruction afterwards, such order is not essential. However, the order is essential during the angle selection stage, i.e., the inference stage of the trained RL policy. This is because during inference, we do not know the structure of the object to be imaged a priori. Thus, it is difficult to make a global personalized decision by selecting all angles at once. What the RL policy does is by making smart angle selections step by step (e.g. by first selecting a few angles around the object to acquire a coarse estimation of the global structure and refine it afterwards) so that the final collection of angles is good. A poorly trained policy may make poor sequential angle selection from the beginning which can eventually lead to an inferior collection of angles.

\subsection{Policy Network Architecture}

\begin{figure}[!htp]
    \centering
    \includegraphics[width=0.9\textwidth]{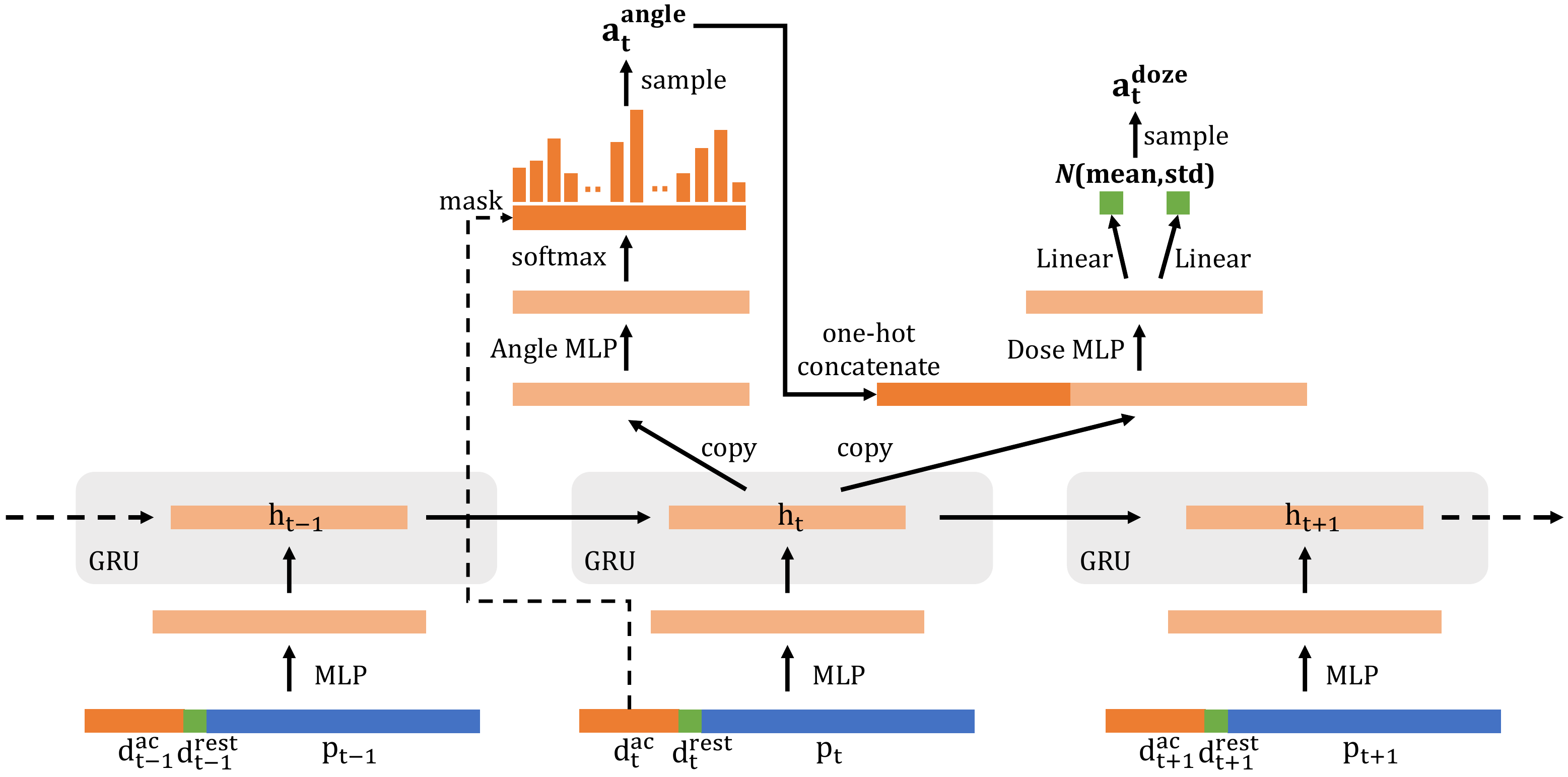}
    \caption{Policy network architecture. Each MLP contains two hidden layers with 512 neurons. We use a multi-layer GRU which contains 3 recurrent layers and each layer has 256 neurons. The Angle MLP has one hidden layer of 512 neurons, and the Dose MLP has 2 hidden layers with 512 neurons.}
    \label{fig:policy_net}
\end{figure}

Because we include all the previous measurements in the state, the dimension of the state vector increases as we take more measurements. To handle the varying dimensionality of the state vector, we represent the policy network as a Recurrent Neural Network (RNN), so all the information from the past measurements can be encoded in the hidden state of the RNN. Specifically, we use the Gated Recurrent Unit (GRU). Besides, the policy network needs to output two different actions: the discrete action for choosing the angle $a_{t}^{\text{angle}}$, and the continuous action for choosing the dose $a_t^{\text{dose}}$. To handle this, we design a special architecture for the policy network, as shown in Figure \ref{fig:policy_net}. We use separate Multi-Layer Perceptron (MLP) after the RNN hidden states for these two actions. $a_t^{\text{angle}}$ is sampled from a probability vector of length $N$, where the value at each entry represents the probability of choosing that angle. We use softmax after the final linear layer to obtain the probability vector. We also introduce a mask to remove the previously chosen angles, which is obtained from $d_t^{\text{ac}}$ that records the used dose distribution on all angles up to time $t$. For the dose usage $a_t^{\text{dose}}$, we assume $a_t^{\text{dose}} \sim \mathcal{N}(\mu^{\text{dose}}, \sigma^{\text{dose}})$, with the mean and std both learned by a MLP. It is natural to determine the amount of dose after the angle is chosen, i.e., $\pi_{\theta}(a|\vec{s})=\pi_{\theta}(a^{\text{dose}},a^{\text{angle}}|\vec{s})=\pi_{\theta}(a^{\text{angle}}|\vec{s})\pi_\theta(a^{\text{dose}}|\vec{s},a^{\text{angle}})$, so we concatenate the one-hot vector of the chosen angle as part of the input for the dose MLP.

\section{Experiments}

\subsection{Experiment Setup}

We train the RL policy on 250 CT images of size $512 \times 512$ from the AAPM dataset of the "2016 NIH-AAPM-Mayo Clinic Low Dose CT Grand Challenge"~\cite{mccollough2016tu}. The 250 training images are slices of the 3D CT image of one patient. The ground-truth images are obtained by using the commercial reconstruction algorithm with normal dose. We use ASTRA Toolbox~\cite{astratool1,astratool2} to generate the sinogram data.
During training, we use SART as the reconstruction algorithm for computing the reward. The possible angles are all integers in [0\degree,360\degree). We use Adam~\cite{adam} to optimize both the policy network and the value network, with a learning rate of $0.0004$, and $\beta_1, \beta_2 = (0.5, 0.999)$. More detailed hyperparameters for PPO and network architecture can be found in the code which will be released upon acceptance of this paper. 
 
After training, we test the learned RL policy on  350 slices of CT images of another patient from the AAPM dataset. We compare the following three scanning strategies: (1) \textbf{UF-AEC}, which selects angles uniformly and distributes the doze by automatic exposure control (AEC)\cite{AEC_1,AEC_2}; (2) \textbf{DS-ED}, which selects angle by a dynamic sample strategy based on entropy from \cite{entropymeth} , while distributes the doze uniformly; (3) \textbf{RL-AD}: which uses the learned personalized policy for both angle selection and dose allocation at each chosen angle.  During testing, we use four different reconstruction algorithms: SART, TV regularization (TV) \cite{rudin1992nonlinear,goldstein2009split}, wavelet frame (WF) regularization \cite{ron1997affine,cai2009split}, and the recently proposed deep learning method PD-net \cite{pdhgnet}.
Note that during the angle selection stage, i.e. the inference stage of the trained RL policy, we do not need to conduct image reconstruction. The decision on the next angle and the associated dose is determined from the sinogram formed by the angles and doses that have already been selected by the RL policy. We run the reconstruction algorithm after all angles are selected by the Rl policy. In other words, the angles/dose selection and image reconstruction are two independent stages. The evaluation metric is PSNR and the structure similarity metric (SSIM) of the reconstructed images. In the testing phase, we add an additional Poisson noise to the sinogram. Recall that the original noise in the sinogram is Gaussian as described in Section 2.3. Note that we add additional noise to test the robustness and generalization ability of our trained RL policy, and we choose Poisson noise since it is widely accepted in CT imaging that the measurement noise is Poisson. The incident photon intensities of the additional Poisson noise are $10^7$, $10^6$ and $10^5$. We denote these three levels of noise as Noise 1, Noise 2 and Noise 3. 

A difficulty in conducting a fair comparison of UF-AEC and DS-ED with RL-AD is that the number of selected angles of RL-AD is personalized and hence different for different subjects (see Figure \ref{fig:hist}). In our experiments below, we choose the number of measurement angles for RD-AEC and DS-ED to be 60, which is the mean number of measurement angles selected by RL-AD over all the 350 test images. Thus, the dose on each measurement angle of UF-AEC and DS-ED is $1/60$. We also note that the deep reconstruction model PD-net is trained from scratch on the 250 images in the training set using 60 angles and cone-beams geometry. 


    

\subsection{Results}
\begin{table}[!htp]
    \centering
    \begin{tabular}{|c|c|c|c|c|}
    \hline
         \multicolumn{2}{|c|}{Reconstruction Method}&RL-AD&DS-ED&UF-AEC\\
         \hline
         \multicolumn{5}{|c|}{Noise 1}\\
         \hline
         \multirow{2}{*}{SART}&PSNR&\textbf{23.48(0.47)}&23.30(0.64)&23.01(0.64)\\
         \cline{2-5}
         &SSIM&\textbf{0.424(0.020)}&0.403(0.023)&0.391(0.022)\\
         \hline
         
         \multirow{2}{*}{TV}&PSNR&\textbf{23.85(0.42)}&23.75(0.41)&23.63(0.38)\\
         \cline{2-5}
         &SSIM&\textbf{0.582(0.030)}&0.579(0.030)&0.578(0.028)\\
         \hline
         \multirow{2}{*}{WF}&PSNR&\textbf{25.14(0.40)}&25.05(0.42)&24.91(0.39)\\
         \cline{2-5}
         &SSIM&\textbf{0.659(0.027)}&0.652(0.027)&0.649(0.026)\\
         \hline
         \multirow{2}{*}{PD-net}&PSNR&\textbf{30.87(0.64)}&30.44(0.51)&30.23(0.46)\\
         \cline{2-5}
         &SSIM&\textbf{0.776(0.036)}&0.771(0.029)&0.773(0.028)\\
         \hline
         \multicolumn{5}{|c|}{Noise 2}\\
         \hline
         \multirow{2}{*}{SART}&PSNR&\textbf{23.15(0.48)}&22.91(0.53)&22.60(0.64)\\
         \cline{2-5}
         &SSIM&\textbf{0.413(0.020)}&0.390(0.024)&0.378(0.024)\\
         \hline
         \multirow{2}{*}{TV}&PSNR&\textbf{23.74(0.40)}&23.50(0.36)&23.27(0.40)\\
         \cline{2-5}
         &SSIM&\textbf{0.580(0.030)}&0.576(0.030)&0.573(0.028)\\
         \hline
         \multirow{2}{*}{WF}&PSNR&\textbf{24.98(0.29)}&24.84(0.41)&24.68(0.39)\\
         \cline{2-5}
         &SSIM&\textbf{0.657(0.027)}&0.649(0.026)&0.646(0.026)\\
         \hline
         \multirow{2}{*}{PD-net}&PSNR&\textbf{30.78(0.64)}&30.35(0.51)&30.15(0.77)\\
         \cline{2-5}
         &SSIM&\textbf{0.774(0.037)}&0.769(0.030)&0.771(0.029)\\
         \hline
         \multicolumn{5}{|c|}{Noise 3}\\
         \hline
        \multirow{2}{*}{SART}&PSNR&\textbf{20.71(0.55)}&20.26(0.72)&19.83(0.66)\\
         \cline{2-5}
         &SSIM&\textbf{0.334(0.026)}&0.304(0.030)&0.291(0.029)\\
         \hline
         \multirow{2}{*}{TV}&PSNR&\textbf{21.73(0.57)}&21.43(0.48)&21.08(0.47)\\
         \cline{2-5}
         &SSIM&\textbf{0.568(0.027)}&0.555(0.026)&0.545(0.026)\\
         \hline
         \multirow{2}{*}{WF}&PSNR&\textbf{23.35(0.48)}&23.05(0.51)&22.72(0.55)\\
         \cline{2-5}
         &SSIM&\textbf{0.636(0.0326)}&0.616(0.027)&0.605(0.028)\\
         \hline
         \multirow{2}{*}{PD-net}&PSNR&\textbf{29.97(0.66)}&29.56(0.51)&29.36(0.47)\\
         \cline{2-5}
         &SSIM&\textbf{0.753(0.038)}&0.746(0.032)&0.747(0.031)\\
         \hline
         \multicolumn{2}{|c|}{Inference Time (s)}&0.46(0.02)&0.21(0.008)&0.20(0.001)\\
         \hline
    \end{tabular}
    
    \caption{This table presents comparisons of different scanning strategies (1-3rd column for RL-AD, DS-ED and UF-AEC respectively) combined with different image reconstruction methods (1-4th row for SART, TV, WF and PD-net respectively).
    Last row presents the inference times of angle selection (in seconds)  of the three compared scanning strategies. The mean (std) of the PSNR and SSIM of the reconstructed images and the inference times are computed among all 350 testing CT images. The best results among the compared algorithms are shown in bold numbers.}
    \label{tab:my_label}
\end{table}

\begin{figure}
    \centering
    \begin{tabular}{c}
    \includegraphics[width=0.8\textwidth]{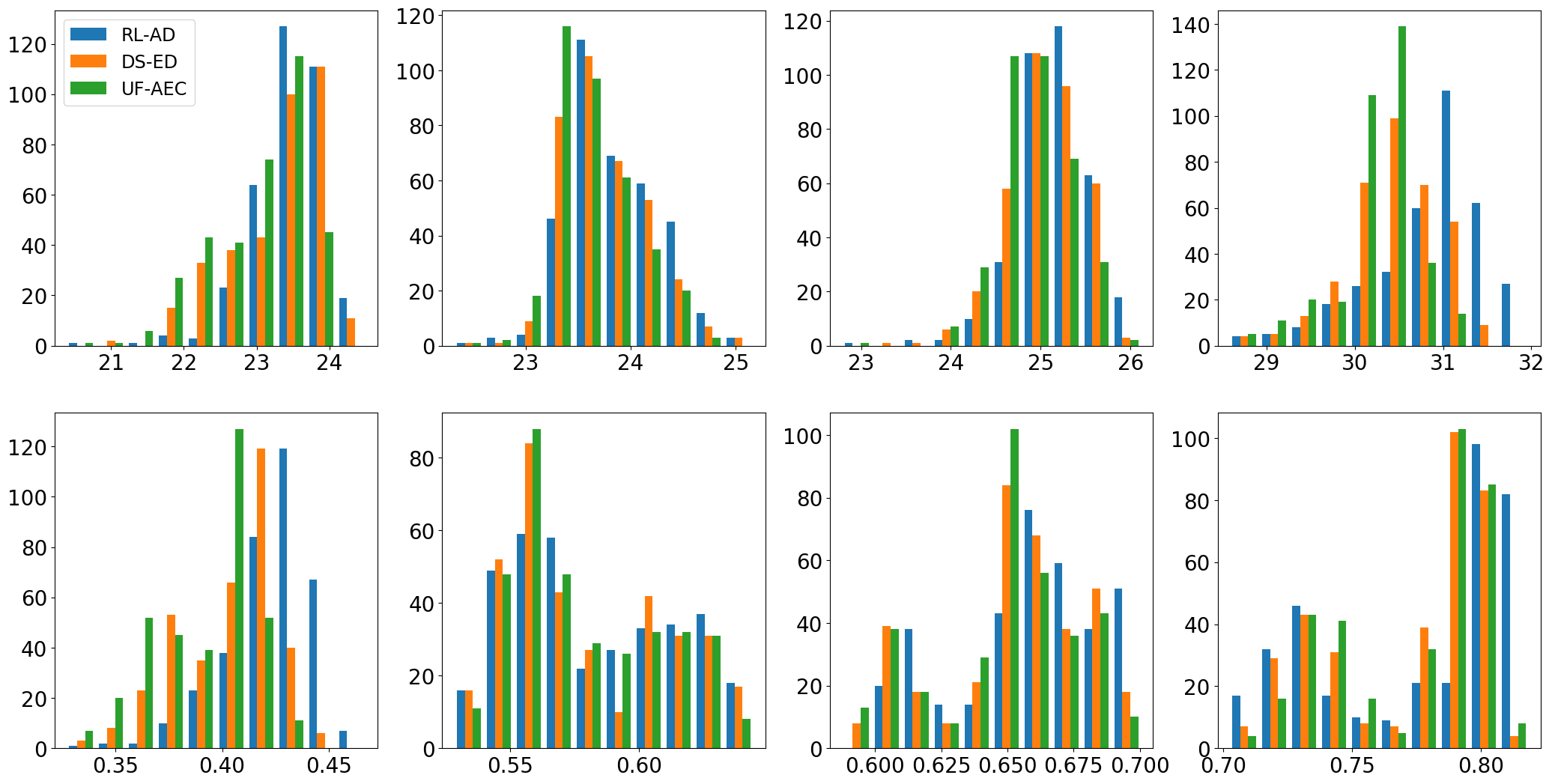}\\
    (a) \\
    \includegraphics[width=0.8\textwidth]{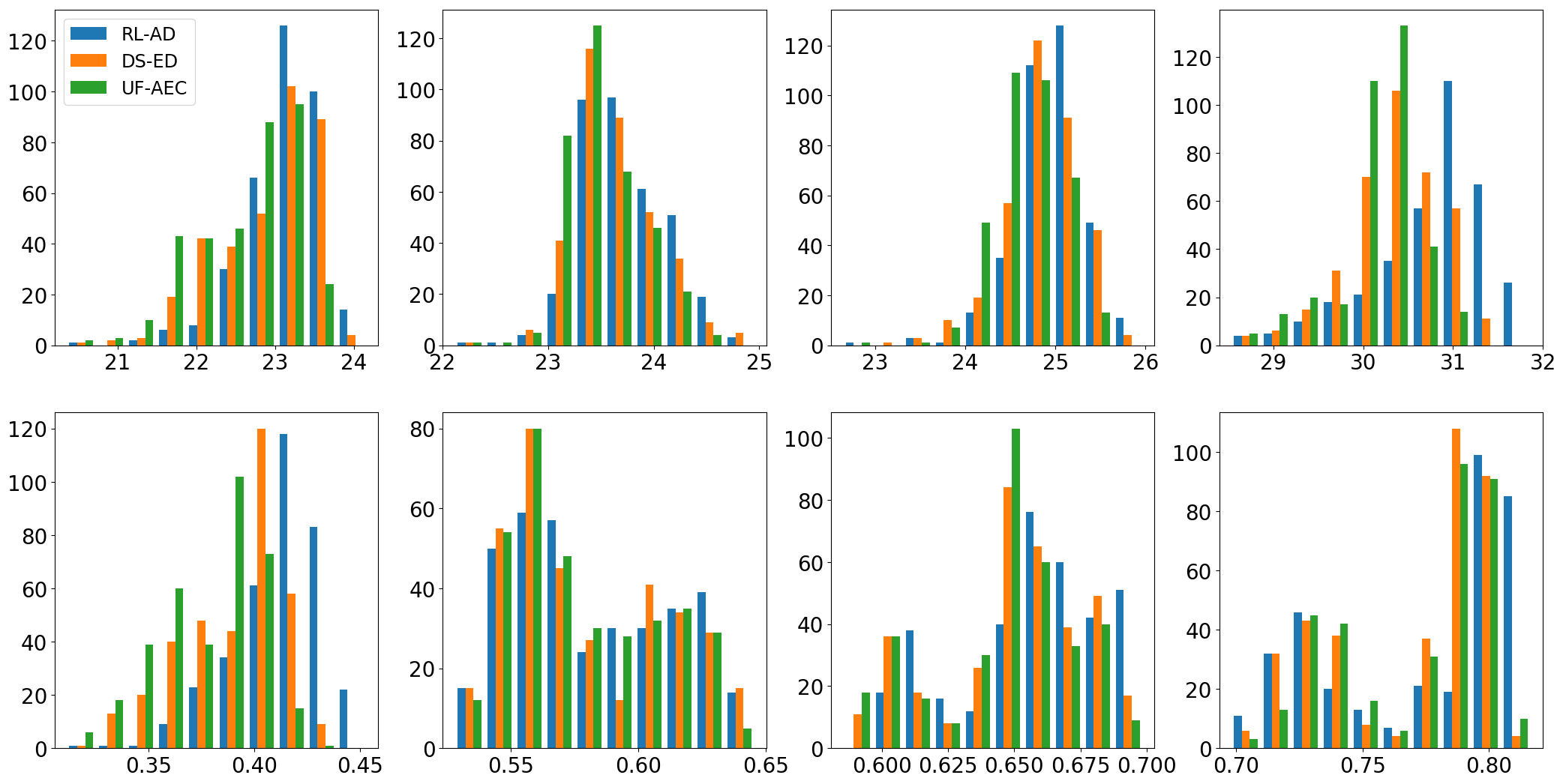}\\
    (b)\\
    \includegraphics[width=0.8\textwidth]{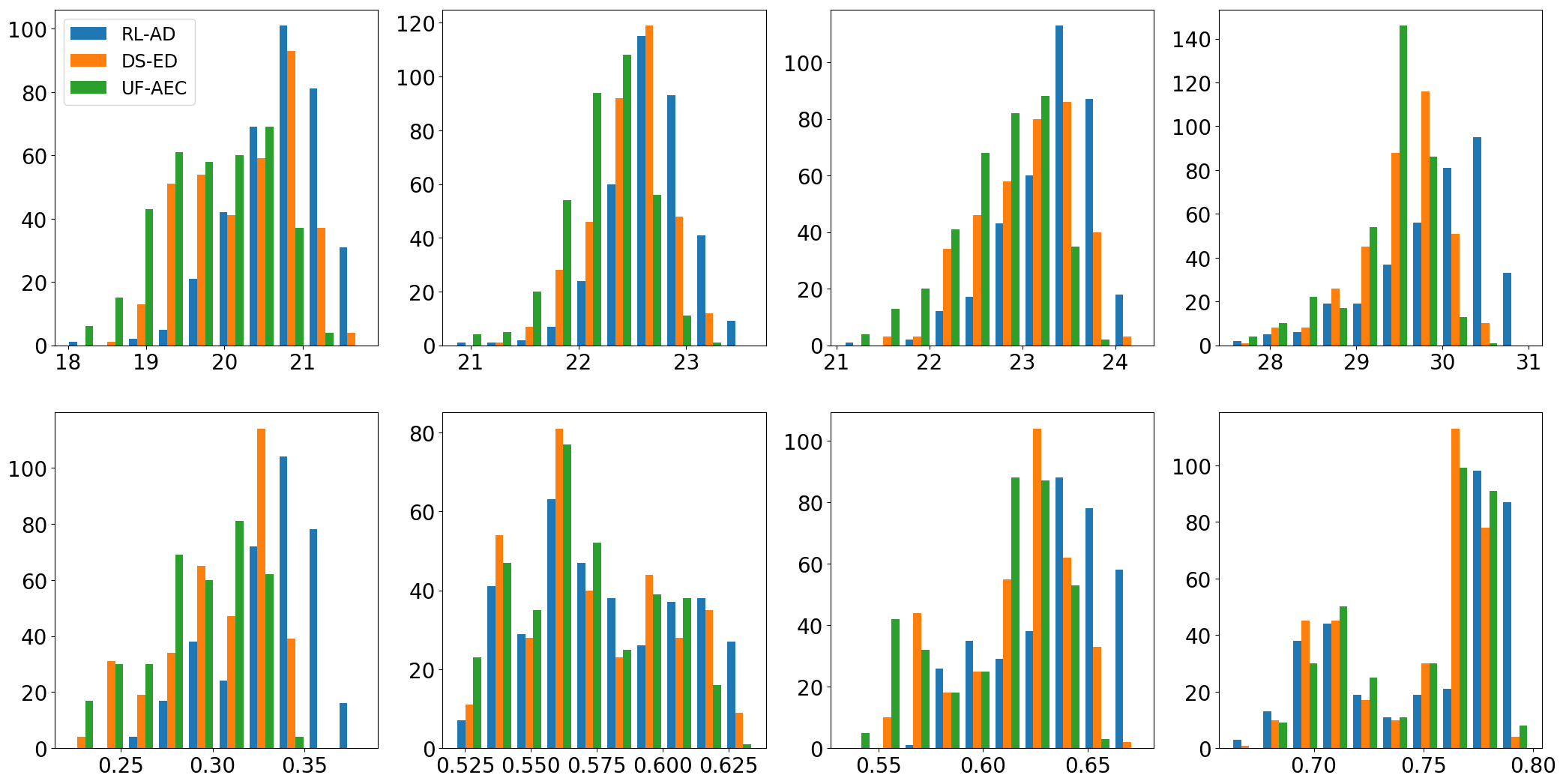}\\
    (c)\\
    \end{tabular}
    \caption{Histograms of PSNR and SSIM from all the 350 test images as shown in Table \ref{tab:my_label}. Figures in (a), (b) and (c) correspond to the three different noise levels. For each (a), (b) and (c), the first row is PSNR and the second is SSIM. Figures from left to right are results from reconstruction methods SART, TV, WF and PD-net respectively. Every sub-figure contains histograms of three scanning strategies, i.e., RL-AD, DS-ED and UF-AEC.}
    \label{fig:psnr_ssim_hist}
\end{figure}

\begin{figure}[htbp]
\centering
\subfigure[]{
    \begin{minipage}[t]{0.5\linewidth}
        \centering
         \includegraphics[width=2.1in]{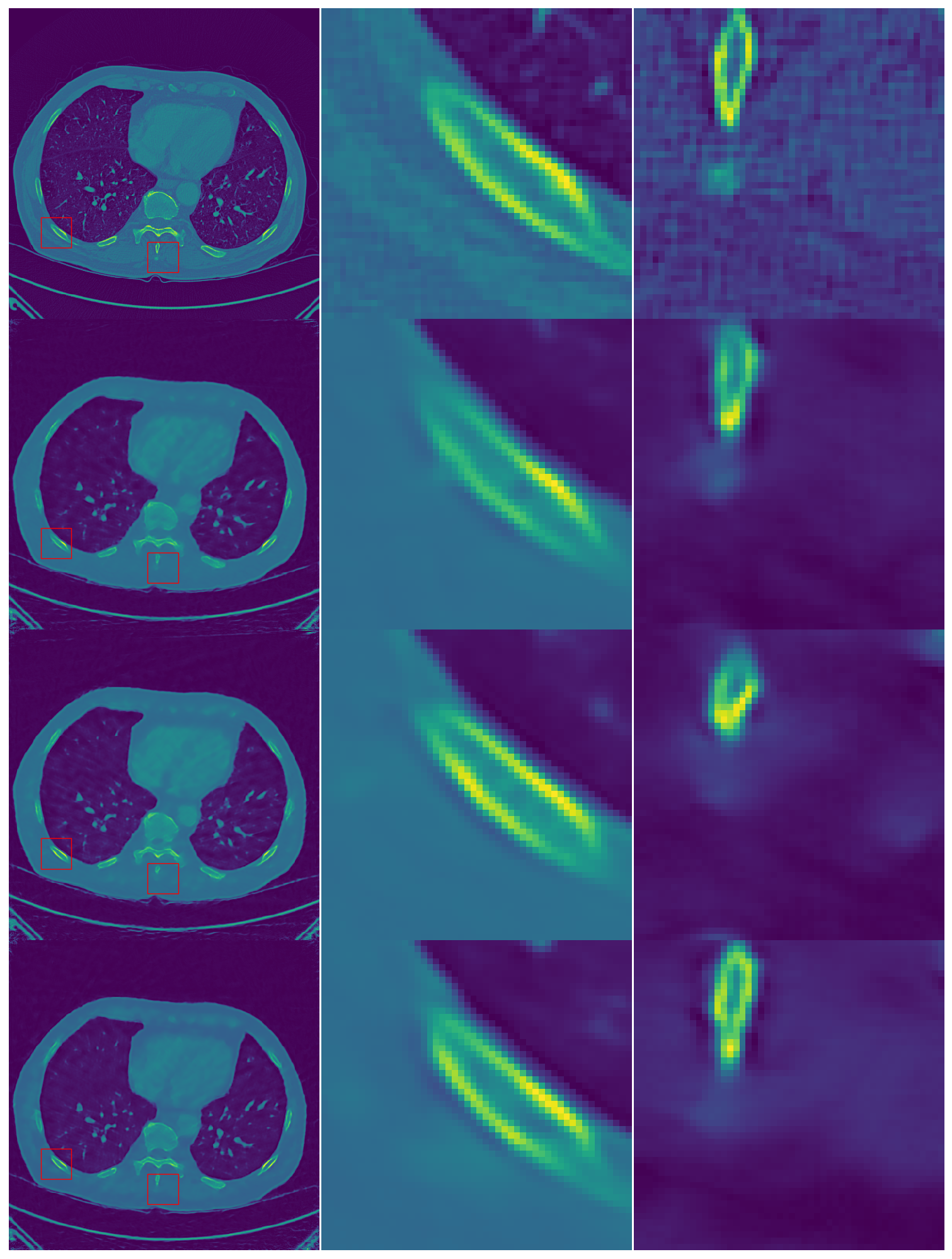}
        \label{fig:imga}
    \end{minipage}
}%
\subfigure[]{
    \begin{minipage}[t]{0.5\linewidth}
        \centering
        \includegraphics[width=2.1in]{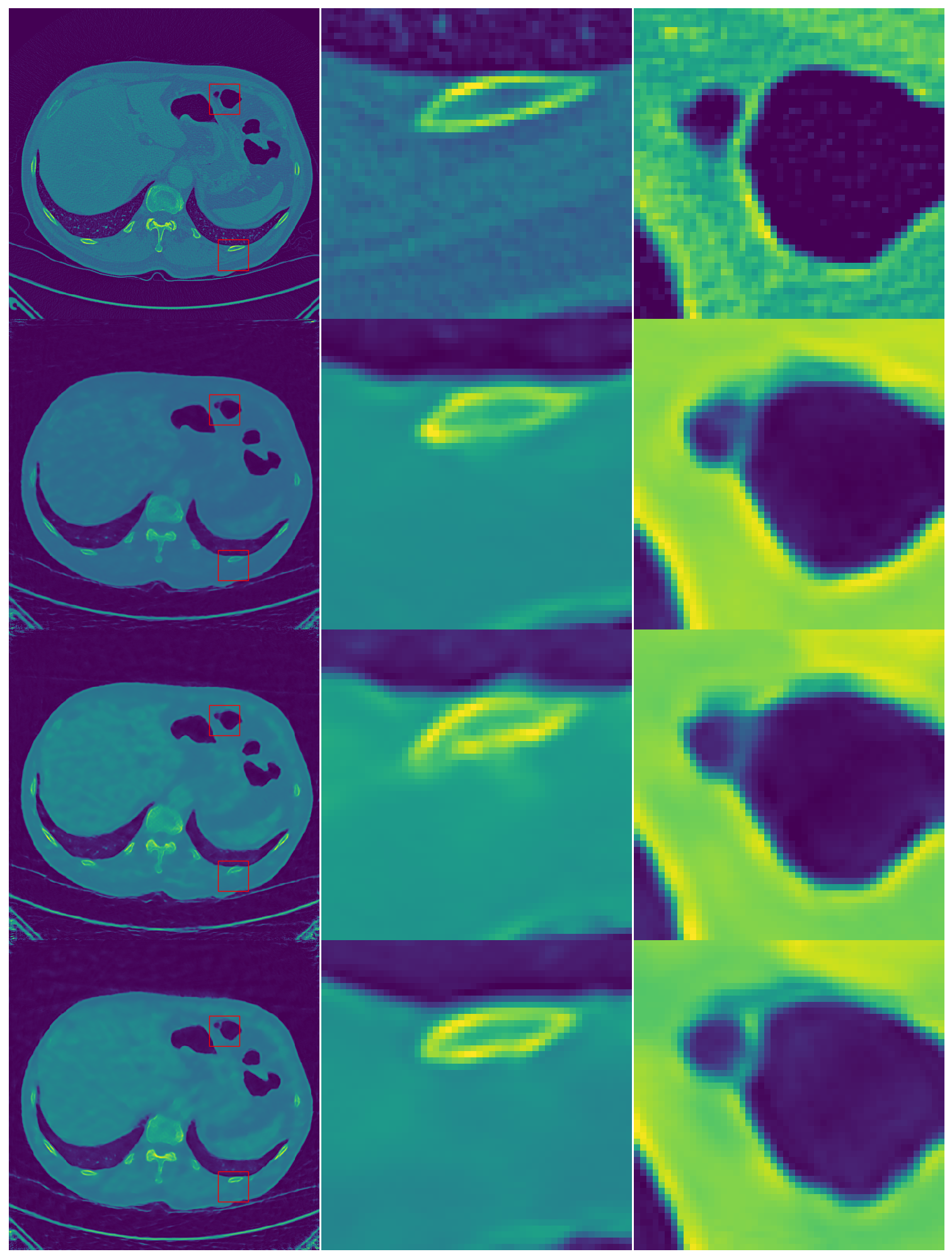}
        \label{fig:imgb}
    \end{minipage}%
}%

\centering
\caption{Two examples of the reconstructed images. The top row contains the ground truth images and their zoom-in views. The second through the fourth row contain results from UF-AEC, DS-ED and RL-AD respectively, and combined with PD-net's reconstruction. Note that RL-AD selects 65 measurement angles for the subject in (a) and 54 measurement angles for the subject in (b).}
\vspace{-0.2cm}
\label{fig:example_reconstruction}
\end{figure}


Table \ref{tab:my_label} presents the mean and standard deviation of the PSNR and SSIM values of the reconstructed images of all compared scanning strategies and reconstruction algorithms. As one can see that the proposed scanning strategy RL-AD significantly outperforms dynamic sampling (DS-ED) and uniform sampling (UF-AEC), while DS-ED outperforms UF-AEC. 


We also note that the RL-policy is trained only using the SART for computing the reward function, whereas the learned policy can generalize well to three other reconstruction algorithms, i.e., the TV regularization, the wavelet frame regularization and the deep learning model PD-net, where it still brings a notable improvement upon the dynamic sampling and uniform scanning baseline in reconstruction quality. We further note that during training, only Gaussian noise was included following the formula given in Section 2.3. The results in Table \ref{tab:my_label} also shows that the trained RL-policy is also transferable to different noise levels.

More fine-grained demonstrations of the results shown in Table \ref{tab:my_label} are given in Figure \ref{fig:psnr_ssim_hist}, where we present the histograms of the compared scanning strategies and reconstructions methods. As one can see that the proposed scanning strategy by RL generally shifted the histogram towards the right and outperforms DS and UF by a significant margin. 


In Figure \ref{fig:example_reconstruction}, we further show two examples of the reconstructed images using the uniform sampling (UF-AEC), the dynamic sampling strategy (DS-ED) and the learned personalized policy (RL-AD),  reconstructed using the deep learning model PD-net. We can see that the reconstructions using the RL policy are of higher qualities than those using random and dynamic sampling strategy, especially from the zoom-in views of the figures.

\begin{figure}
    \centering
    \begin{tabular}{ccc}
    \includegraphics[width=.23\textwidth]{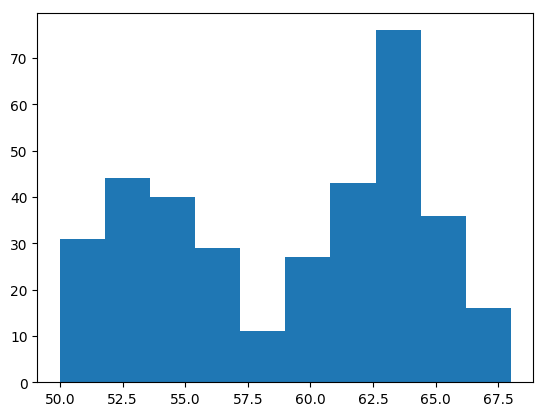}&   \includegraphics[width=.23\textwidth]{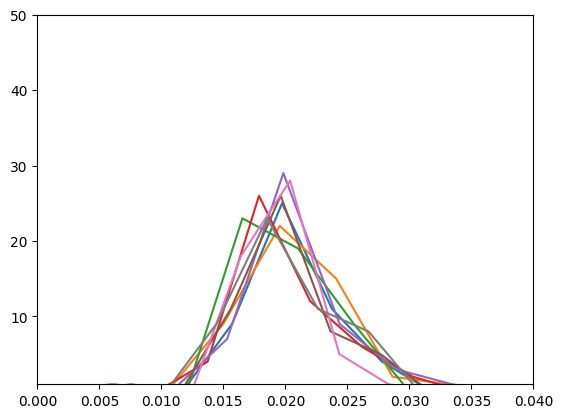}& 
    \includegraphics[width=.23\textwidth]{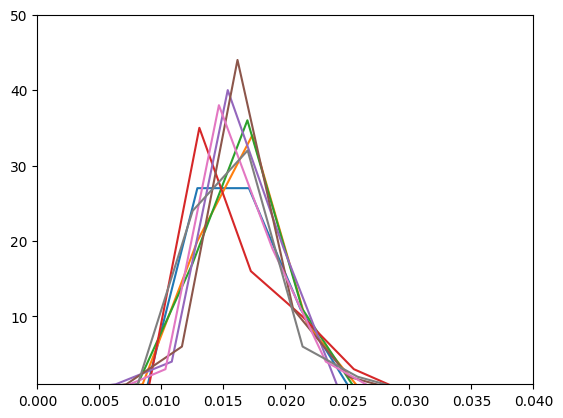}
    \\
    (a)  & (b) & (c)
    \end{tabular}
    \caption{(a) Distribution of number of measurements of the learned policy (RL-AD) on all the 350 testing CT images. (b) Dose usage distribution of 8 images that use around 54 measurements. (c) Dose usage distribution of 8 images that use around 65 measurements.   
    }
    \label{fig:hist}
\end{figure}
\begin{figure}[htbp]
\centering

\subfigure[]{
    \begin{minipage}[t]{0.35\linewidth}
        \centering
        \includegraphics[width=1.7in]{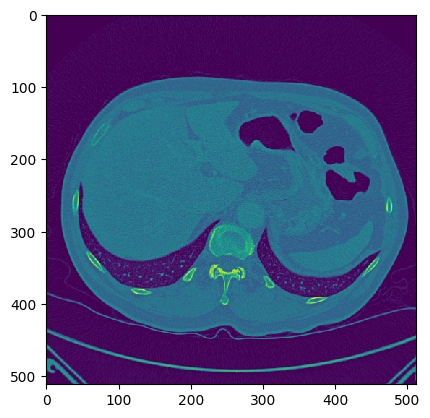}
        \label{fig:imgb}
    \end{minipage}%
}%
 \subfigure[]{
    \begin{minipage}[t]{0.35\linewidth}
        \centering
        \includegraphics[width=1.7in]{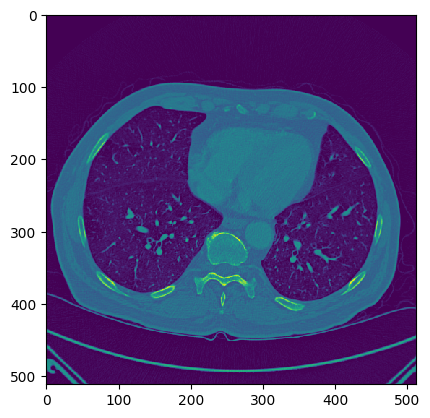}
        \label{fig:imgd}
    \end{minipage}%
}%
\centering
\caption{(a): an example image that takes 54 measurements. (b): an example image that takes 64 measurements. We can see that the images for which RL selects more measurement angles contains more structures. }
\vspace{-0.2cm}
\label{fig:compare_fig}
\end{figure}
\begin{figure}[!htp]
    \centering
    \begin{tabular}{cc}
    \includegraphics[width=.4\textwidth]{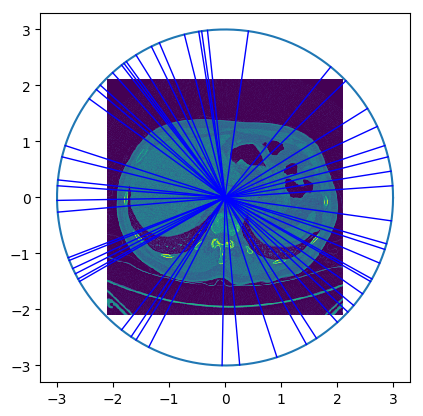}&  
    \includegraphics[width=.4\textwidth]{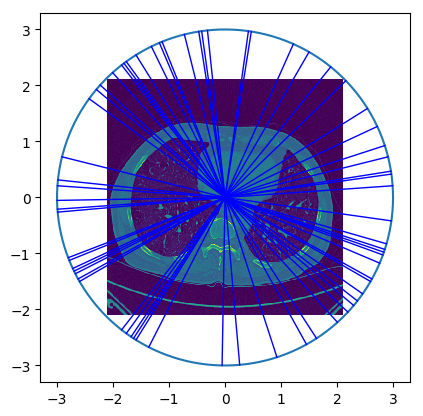}\\
    (a) & (b) \\
    \includegraphics[width=.4\textwidth]{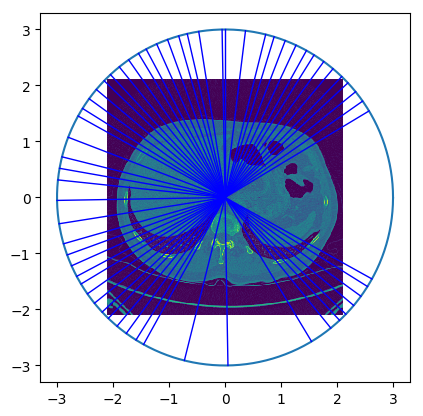}& 
    \includegraphics[width=.4\textwidth]{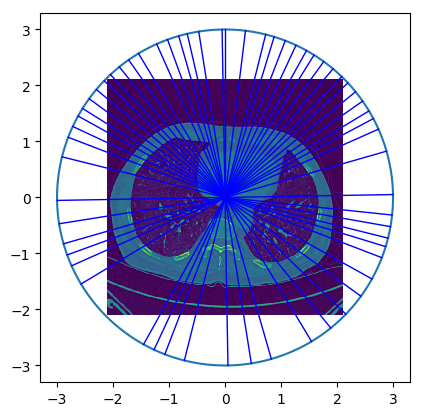}\\
    (c) & (d)\\
    \end{tabular}
    \caption{The angle selection of the CT image in Figure \ref{fig:compare_fig}. Top row: RL-AD, bottom row: DS-ED. The lines show the selected angles.
    }
    \label{fig:angle-dose-circle}
\end{figure}

We plot the distribution of number of measurements taken by the learned personalized policy (RL-AD) in Figure \ref{fig:hist} (a). The result demonstrates that for different subjects, the learned RL policy selects different number of angles and dose allocations. In Figure \ref{fig:hist} (b) and (c), we take 8 images on which the learned RL policy selects 54 and 64 angles respectively and plot the distributions of the dose usage of these images. It can be seen that images using the same number of measurement angles have very similar dose allocations, and images that have more measurement angles use less dose at each angle. In Figure \ref{fig:compare_fig}, we show 2 example images where the RL policy selects 54 and 65 measurement angles respectively. We can see that images upon which the RL policy selects more measurement angles have more structures in the image, and thus more information/measurements need to be collected to obtain a high-quality reconstruction. 
In Figure \ref{fig:angle-dose-circle}, we present the selected angles and part of the dose allocation on the subjects shown in Figure \ref{fig:compare_fig}.



\section{Conclusion}
In this paper,  we proposed to use reinforcement learning to learn a personalized CT scanning strategy for measurement angle selection and dose allocation. We formulated the CT scanning process as a Markov Decision Process, and used the PPO algorithm to solve it. After training on 250 real 2D CT images, we validated the learned personalized scanning policy on another 350 CT images. Our validation showed that the personalized scanning policy lead to better overall reconstruction results in terms of PSNR values, and generalized well to be combined with different reconstruction algorithms. We also demonstrated that the personalized policy can indeed adjust its angle selection and dose allocations adaptively to different subjects. One drawback of the proposed method is the long training time (approximately 24 hours) even for 2D images, because RL algorithms usually need lots of simulation samples to converge, and to compute the reward in our formulated MDP requires running a reconstruction algorithm at each time step. This might prohibits the application of our method to 3D cases.

\section*{Acknowledgements}
Bin Dong is supported in part by National Natural Science Foundation of
China (NSFC) grant No. 11831002, Beijing Natural Science Foundation (No. 180001) and Beijing Academy of Artificial Intelligence (BAAI). Xu Yang was partially supported by the NSF grant DMS-1818592.

\bibliographystyle{unsrt}
\bibliography{CT}

\begin{thebibliography}{10}

\bibitem{FBP}
Alexander Katsevich.
\newblock Theoretically exact filtered backprojection-type inversion algorithm
  for spiral ct.
\newblock {\em Siam Journal on Applied Mathematics}, 62(6):2012--2026, 2002.

\bibitem{ART}
R.~Gordon, R.~Benderab, and G.~T. Herman.
\newblock Algebraic reconstruction techniques (art) for three-dimensional
  electron microscopy and x-ray photography.
\newblock {\em Journal of Theoretical Biology}, 1970.

\bibitem{compress_sense}
D.~Donoho.
\newblock Compressed sensing.
\newblock {\em IEEE Transactions on Information Theory}, 52(4):1289--1306,
  2006.

\bibitem{drip}
Emmanuel~J. Candes, Yonina~C. Eldar, et~al.
\newblock Compressed sensing with coherent and redundant dictionaries.
\newblock 2010.

\bibitem{admm1}
T.~Goldstein and S.~Osher.
\newblock The split bregman method for l1-regularized problems.
\newblock {\em SIAM Journal Imaging Sciences}, 2(2):323--343, 2009.

\bibitem{admm2}
S.~Boyd, N.~Parikh, et~al.
\newblock Distributed optimization and statistical learning via the alternating
  direction method of multipliers.
\newblock {\em Foundations and Trends in Machine Learning}, 3(1):1--122, 2011.

\bibitem{cai2009split}
Jian-Feng Cai, Stanley Osher, and Zuowei Shen.
\newblock Split bregman methods and frame based image restoration.
\newblock {\em Multiscale Modeling and Simulation}, 8(2):337--369, 2009.

\bibitem{pdhg1}
Antonin Chambolle and Thomas Pock.
\newblock A first-order primal-dual algorithm for convex problems with
  applications to imaging.
\newblock {\em Journal of Mathematical Imaging and Vision}, 40(1):120--145,
  2011.

\bibitem{pdhg2}
Mingqiang Zhu and Tony Chan.
\newblock An efficient primal-dual hybrid gradient algorithm for total
  variation image restoration.
\newblock {\em UCLA CAM Report}, 2008.

\bibitem{pdhg3}
Ernie Esser, Xiaoqun Zhang, et~al.
\newblock A general framework for a class of first order primal-dual algorithms
  for convex optimization in imaging science.
\newblock {\em SIAM Journal on Imaging Sciences}, 3(4):1015--1046, 22010.

\bibitem{rudin1992nonlinear}
Leonid~I. Rudin, Stanley Osher, and Emad Fatemi.
\newblock Nonlinear total variation based noise removal algorithms.
\newblock {\em Physical D: nonlinear phenomena}, 60(1-4):259--268, 1992.

\bibitem{buades2005non}
Antoni Buades, Bartomeu Coll, and J-M Morel.
\newblock A non-local algorithm for image denoising.
\newblock In {\em 2005 IEEE Computer Society Conference on Computer Vision and
  Pattern Recognition (CVPR'05)}, volume~2, pages 60--65. IEEE, 2005.

\bibitem{dabov2007image}
Kostadin Dabov, Alessandro Foi, Vladimir Katkovnik, and Karen Egiazarian.
\newblock Image denoising by sparse 3-d transform-domain collaborative
  filtering.
\newblock {\em IEEE Transactions on image processing}, 16(8):2080--2095, 2007.

\bibitem{gu2014weighted}
Shuhang Gu, Lei Zhang, Wangmeng Zuo, and Xiangchu Feng.
\newblock Weighted nuclear norm minimization with application to image
  denoising.
\newblock In {\em Proceedings of the IEEE conference on computer vision and
  pattern recognition}, pages 2862--2869, 2014.

\bibitem{daubechies1992ten}
Ingrid Daubechies.
\newblock {\em Ten lectures on wavelets}, volume~61.
\newblock Siam, 1992.

\bibitem{mallat1999wavelet}
St{\'e}phane Mallat.
\newblock {\em A wavelet tour of signal processing}.
\newblock Elsevier, 1999.

\bibitem{dong2010mra}
Bin Dong, Zuowei Shen, et~al.
\newblock Mra based wavelet frames and applications.
\newblock {\em IAS Lecture Notes Series, Summer Program on ``The Mathematics of
  Image Processing'', Park City Mathematics Institute}, 19, 2010.

\bibitem{elad2006image}
Michael Elad and Michal Aharon.
\newblock Image denoising via sparse and redundant representations over learned
  dictionaries.
\newblock {\em IEEE Transactions on Image processing}, 15(12):3736--3745, 2006.

\bibitem{cai2014data}
Jian-Feng Cai, Hui Ji, Zuowei Shen, and Gui-Bo Ye.
\newblock Data-driven tight frame construction and image denoising.
\newblock {\em Applied and Computational Harmonic Analysis}, 37(1):89--105,
  2014.

\bibitem{tai2016multiscale}
Cheng Tai and E~Weinan.
\newblock Multiscale adaptive representation of signals: I. the basic
  framework.
\newblock {\em The Journal of Machine Learning Research}, 17(1):4875--4912,
  2016.

\bibitem{osher2017low}
Stanley Osher, Zuoqiang Shi, and Wei Zhu.
\newblock Low dimensional manifold model for image processing.
\newblock {\em SIAM Journal on Imaging Sciences}, 10(4):1669--1690, 2017.

\bibitem{wang2016perspective}
Ge~Wang.
\newblock A perspective on deep imaging.
\newblock {\em IEEE Access}, 4:8914--8924, 2016.

\bibitem{wang2017machine}
Ge~Wang, Mannudeep Kalra, and Colin~G Orton.
\newblock Machine learning will transform radiology significantly within the
  next 5 years.
\newblock {\em Medical physics}, 44(6):2041--2044, 2017.

\bibitem{mccann2017convolutional}
Michael~T McCann, Kyong~Hwan Jin, and Michael Unser.
\newblock Convolutional neural networks for inverse problems in imaging: A
  review.
\newblock {\em IEEE Signal Processing Magazine}, 34(6):85--95, 2017.

\bibitem{wang2018image}
Ge~Wang, Jong~Chu Ye, Klaus Mueller, and Jeffrey~A Fessler.
\newblock Image reconstruction is a new frontier of machine learning.
\newblock {\em IEEE transactions on medical imaging}, 37(6):1289--1296, 2018.

\bibitem{zhang2020review}
Hai-Miao Zhang and Bin Dong.
\newblock A review on deep learning in medical image reconstruction.
\newblock {\em Journal of the Operations Research Society of China}, pages
  1--30, 2020.

\bibitem{chen2017low}
Hu~Chen, Yi~Zhang, Mannudeep~K Kalra, Feng Lin, Yang Chen, Peixi Liao, Jiliu
  Zhou, and Ge~Wang.
\newblock Low-dose ct with a residual encoder-decoder convolutional neural
  network.
\newblock {\em IEEE transactions on medical imaging}, 36(12):2524--2535, 2017.

\bibitem{jin2017deep}
Kyong~Hwan Jin, Michael~T McCann, Emmanuel Froustey, and Michael Unser.
\newblock Deep convolutional neural network for inverse problems in imaging.
\newblock {\em IEEE Transactions on Image Processing}, 26(9):4509--4522, 2017.

\bibitem{kang2017deep}
Eunhee Kang, Junhong Min, and Jong~Chul Ye.
\newblock A deep convolutional neural network using directional wavelets for
  low-dose x-ray ct reconstruction.
\newblock {\em Medical physics}, 44(10):e360--e375, 2017.

\bibitem{zhang2018sparse}
Zhicheng Zhang, Xiaokun Liang, Xu~Dong, Yaoqin Xie, and Guohua Cao.
\newblock A sparse-view ct reconstruction method based on combination of
  densenet and deconvolution.
\newblock {\em IEEE transactions on medical imaging}, 37(6):1407--1417, 2018.

\bibitem{yang2018low}
Qingsong Yang, Pingkun Yan, Yanbo Zhang, Hengyong Yu, Yongyi Shi, Xuanqin Mou,
  Mannudeep~K Kalra, Yi~Zhang, Ling Sun, and Ge~Wang.
\newblock Low-dose ct image denoising using a generative adversarial network
  with wasserstein distance and perceptual loss.
\newblock {\em IEEE transactions on medical imaging}, 37(6):1348--1357, 2018.

\bibitem{shen2018intelligent}
Chenyang Shen, Yesenia Gonzalez, Liyuan Chen, Steve~B Jiang, and Xun Jia.
\newblock Intelligent parameter tuning in optimization-based iterative ct
  reconstruction via deep reinforcement learning.
\newblock {\em IEEE transactions on medical imaging}, 37(6):1430--1439, 2018.

\bibitem{rip}
E.~J. Candes, J.~Romberg, and T.~Tao.
\newblock Robust uncertainty principles: Exact signal reconstruction from
  highly incomplete fourier information.
\newblock {\em IEEE Transactions on Information Theory}, 52(2), 2006.

\bibitem{settles2009active}
Burr Settles.
\newblock Active learning literature survey.
\newblock Technical report, University of Wisconsin-Madison Department of
  Computer Sciences, 2009.

\bibitem{RLoverview}
Yuxi Li.
\newblock Deep reinforcement learning: An overview.
\newblock {\em arXiv:1701.07274}, 2017.

\bibitem{zhang2006multi}
J~Zhang, G~Yang, YZ~Lee, Y~Cheng, B~Gao, Q~Qiu, JP~Lu, and O~Zhou.
\newblock A multi-beam x-ray imaging system based on carbon nanotube field
  emitters.
\newblock In {\em Medical Imaging 2006: Physics of Medical Imaging}, volume
  6142, page 614204. International Society for Optics and Photonics, 2006.

\bibitem{multisource}
Ge~Wang and Hengyong Yu.
\newblock A scheme for multisource interior tomography.
\newblock {\em Medical physics}, 36(8):3575--3581, July 2009.

\bibitem{park2020multi}
Junyoung Park, Jaeik Jung, Amar~Prasad Gupta, Jeongtae Soh, Changwon Jeong,
  Jeungsun Ahn, Seungryong Cho, Kwon-Ha Yoon, Dongkeun Kim, Mallory Mativenga,
  et~al.
\newblock Multi-beam x-ray source based on carbon nanotube emitters for
  tomosynthesis system.
\newblock In {\em Medical Imaging 2020: Physics of Medical Imaging}, volume
  11312, page 113122E. International Society for Optics and Photonics, 2020.

\bibitem{low_discrepancy}
Ryutarou Ohbuchi and Masaki Aono.
\newblock Quasi-monte carlo rendering with adaptive sampling.
\newblock 1996.

\bibitem{uniform_sample}
K.A.Mohan, S.V.Venkatakrishnan, E.B.Gulsoy J.W.Gibbs, X.Xiao, M.~D. Graef,
  P.~W. Voorhees, and C.~A. Bouman.
\newblock Timbir: A method for time-space reconstruction from interlaced views.
\newblock {\em IEEE Transactions on Computational Imaging}, pages 96--111,
  2015.

\bibitem{model_sample}
K.~Mueller.
\newblock Selection of optimal views for computed tomography reconstruction.
\newblock {\em Patent WO}, Jan. 28 2011.

\bibitem{model_sample2}
Z.Wang and G.R.Arce.
\newblock Variable density compressed image sampling.
\newblock {\em Image Processing, IEEE Transactions}, 19(1):264--270, 2010.

\bibitem{BCS}
Shihao Ji, Ya~Xue, and Lawrence Carin.
\newblock Bayesian compressive sensing.
\newblock {\em IEEE Transactions on Signal Processing}, 56(6):2346--2356, 2008.

\bibitem{CSBayesian}
Matthias~W Seeger and Hannes Nickisch.
\newblock Compressed sensing and bayesian experimental design.
\newblock In {\em Proceedings of the 25th international conference on Machine
  learning}, pages 912--919, 2008.

\bibitem{inform_gain}
K~Joost Batenburg, Willem~Jan Palenstijn, P{\'e}ter Bal{\'a}zs, and Jan
  Sijbers.
\newblock Dynamic angle selection in binary tomography.
\newblock {\em Computer Vision and Image Understanding}, 117(4):306--318, 2013.

\bibitem{DS_information_gain}
Andrei Dabravolski, Kees~Joost Batenburg, and Jan Sijbers.
\newblock Dynamic angle selection in x-ray computed tomography.
\newblock {\em Nuclear Instruments and Methods in Physics Research Section B:
  Beam Interactions with Materials and Atoms}, 324:17--24, April 2014.

\bibitem{SLAD2}
G.~D. Godaliyadda, M.~A.~Uchic D.~Hye~Ye, M.~A. Groeber, G.~T. Buzzard, and
  C.~A. Bouman.
\newblock A supervised learning approach for dynamic sampling.
\newblock {\em S\&T Imaging. International Society for Optics and Photonics},
  2016.

\bibitem{SLADS}
G.~M. Dilshan~P. Godaliyadda, Dong~Hye Ye, Michael~D. Uchic, Michael~A.
  Groeber, Gregery~T. Buzzard, and Charles~A. Bouman.
\newblock A framework for dynamic image sampling based on supervised learning
  (slads).
\newblock {\em arXiv:1703.04653}, 2017.

\bibitem{SLADS2}
Yan Zhang, G.~M.~Dilshan Godaliyadda, Nicola Ferrier, Emine~B. Gulsoy,
  Charles~A. Bouman, and Charudatta Phatak.
\newblock Slads-net: Supervised learning approach for dynamic sampling using
  deep neural networks.
\newblock {\em Electronic Imaging, Computational Imaging XVI}, 2018.

\bibitem{SLADS3}
Shijie Zhang, Zhengtian Song, GM~Dilshan~P Godaliyadda, Dong~Hye Ye, Azhad~U
  Chowdhury, Atanu Sengupta, Gregery~T Buzzard, Charles~A Bouman, and Garth~J
  Simpson.
\newblock Dynamic sparse sampling for confocal raman microscopy.
\newblock {\em Analytical chemistry}, 90(7):4461--4469, 2018.

\bibitem{SLADS4}
Abderrahim Halimi, Philippe Ciuciu, Aongus Mccarthy, Stephen Mclaughlin, and
  Gerald Buller.
\newblock Fast adaptive scene sampling for single-photon 3d lidar images.
\newblock {\em IEEE CAMSAP 2019 - International Workshop on Computational
  Advances in Multi-Sensor Adaptive Processing}, Dec. 2019.

\bibitem{SLADS5}
Etienne Monier, Nathalie~Brun Thomas~Oberlin, Xiaoyan Li, Marcel Tenc, and
  Nicolas~Dobigeon ~.
\newblock Fast reconstruction of atomic-scale stem-eels images from sparse
  sampling.
\newblock {\em Ultramicroscopy}, 2020.

\bibitem{RLSTEM}
Jeffrey~M. Ede.
\newblock Adaptive partial scanning transmission electron microscopy with
  reinforcement learning.
\newblock {\em arXiv:2004.02786}.

\bibitem{bello2016neural}
Irwan Bello, Hieu Pham, Quoc~V Le, Mohammad Norouzi, and Samy Bengio.
\newblock Neural combinatorial optimization with reinforcement learning.
\newblock {\em arXiv preprint arXiv:1611.09940}, 2016.

\bibitem{kool2018attention}
Wouter Kool, Herke Van~Hoof, and Max Welling.
\newblock Attention, learn to solve routing problems!
\newblock {\em arXiv preprint arXiv:1803.08475}, 2018.

\bibitem{mittal2019learning}
Akash Mittal, Anuj Dhawan, Sahil Manchanda, Sourav Medya, Sayan Ranu, and Ambuj
  Singh.
\newblock Learning heuristics over large graphs via deep reinforcement
  learning.
\newblock {\em arXiv preprint arXiv:1903.03332}, 2019.

\bibitem{auto_explore}
Louis Ly and Yen-Hsi~Richard Tsai.
\newblock Autonomous exploration, reconstruction, and surveillance of 3d
  environments aided by deep learning.
\newblock {\em arXiv:1809.06025}, 2018.

\bibitem{sutton2018reinforcement}
Richard~S Sutton and Andrew~G Barto.
\newblock {\em Reinforcement learning: An introduction}.
\newblock MIT press, 2018.

\bibitem{watkins1992q}
Christopher~JCH Watkins and Peter Dayan.
\newblock Q-learning.
\newblock {\em Machine learning}, 8(3-4):279--292, 1992.

\bibitem{DQN}
Volodymyr Mnih, Koray Kavukcuoglu, and David Silver.
\newblock Human-level control through deep reinforcement learning.
\newblock {\em Nature}, 518, 2015.

\bibitem{mnih2015human}
Volodymyr Mnih, Koray Kavukcuoglu, David Silver, Andrei~A Rusu, Joel Veness,
  Marc~G Bellemare, Alex Graves, Martin Riedmiller, Andreas~K Fidjeland, Georg
  Ostrovski, et~al.
\newblock Human-level control through deep reinforcement learning.
\newblock {\em nature}, 518(7540):529--533, 2015.

\bibitem{sutton2000policy}
Richard~S Sutton, David~A McAllester, Satinder~P Singh, and Yishay Mansour.
\newblock Policy gradient methods for reinforcement learning with function
  approximation.
\newblock In {\em Advances in neural information processing systems}, pages
  1057--1063, 2000.

\bibitem{silver2014deterministic}
David Silver, Guy Lever, Nicolas Heess, Thomas Degris, Daan Wierstra, and
  Martin Riedmiller.
\newblock Deterministic policy gradient algorithms.
\newblock In {\em International conference on machine learning}, pages
  387--395. PMLR, 2014.

\bibitem{ppo}
J.~Schulman and F.~Wolski.
\newblock Proximal policy optimization algorithms.
\newblock {\em arXiv:1707.06347v2}, 2017.

\bibitem{FastCT1985}
Siddon RL.
\newblock Fast calculation of the exact radiological path for a
  three-dimensional ct array.
\newblock {\em Medical Physics}, 2(12):252--5, 1985.

\bibitem{SART}
K.~Mueller, R.~Yagel, and J.J. Wheller.
\newblock Anti-aliased three-dimensional cone-beam reconstruction of
  low-contrast objects with algebraic methods.
\newblock {\em IEEE Transactions On Medical Imaging}, 6(18):519--537, 1999.

\bibitem{TVmodel1}
Emil~Y Sidky and Xiaochuan Pan.
\newblock Image reconstruction in circular cone-beam computed tomography by
  constrained, total-variation minimization.
\newblock {\em Physics in medicine and biology}, 53:4777, 2008.

\bibitem{WaveletFrame}
Bin Dong, Jia Li, and Zuowei Shen.
\newblock X-ray ct image reconstruction via wavelet frame based regularization
  and radon domain inpainting.
\newblock {\em Journal of Scientific Computing}, 54(2-3):333--349, 2013.

\bibitem{CT_sim}
Lifeng Yu, Maria Shiung, Dayna Jondal, and Cynthia~H McCollough.
\newblock Development and validation of a practical lower-dose-simulation tool
  for optimizing computed tomography scan protocols.
\newblock {\em Journal of Computer Assisted Tomography}, 36(4):477--487, July
  2012.

\bibitem{goldstein2009split}
Tom Goldstein and Stanley Osher.
\newblock The split bregman method for $l_1$-regularized problems.
\newblock {\em SIAM Journal on Imaging Sciences}, 2(2):323--343, 2009.

\bibitem{ron1997affine}
Amos Ron and Zuowei Shen.
\newblock Affine systems in {$ L_{2}(\mathbb{R}^{d})$}: The analysis of the
  analysis operator.
\newblock {\em Journal of Functional Analysis}, 148(2):408--447, 1997.

\bibitem{pdhgnet}
Jonas Adler and Ozan Oktem.
\newblock Learned primal-dual reconstruction.
\newblock {\em IEEE Transactions on Medical Imaging}, 37(6):1322--1332, 2018.

\bibitem{mccollough2016tu}
C~McCollough.
\newblock Tu-fg-207a-04: Overview of the low dose ct grand challenge.
\newblock {\em Medical physics}, 43(6Part35):3759--3760, 2016.

\bibitem{astratool1}
W.~van Aarle, W.~J. Palenstijn, J.~Cant, E.~Janssens, F.~Bleichrodt,
  A.~Dabravolski, J.~De Beenhouwer, K.~J. Batenburg, and J.~Sijbers.
\newblock Fast and flexible x-ray tomography using the astra toolbox.
\newblock {\em Optics Express}, 22(24):25129--25147, 2016.

\bibitem{astratool2}
W.~van Aarle, W.~J. Palenstijn, J.~De Beenhouwer, T.~Altantzis, S.~Bals, K.~J.
  Batenburg, and J.~Sijbers.
\newblock The astra toolbox: A platform for advanced algorithm development in
  electron tomography.
\newblock {\em Ultramicroscopy}, (24):35--47, 2015.

\bibitem{adam}
Diederik~P. Kingma and Jimmy Ba.
\newblock Adam: A method for stochastic optimization.
\newblock {\em arXiv:1412.6980}, 2014.

\bibitem{AEC_1}
Michael Gies, Willi~A. Kalender, Heiko Wolf, and Christoph Suess.
\newblock Dose reduction in ct by anatomically adapted tube current modulation.
  i. simulation studies.
\newblock July 1999.

\bibitem{AEC_2}
Willi~A. Kalender, Heiko Wolf, and Christoph Suess.
\newblock Dose reduction in ct by anatomically adapted tube current modulation.
  ii. phantom measurements.
\newblock August 1999.

\bibitem{entropymeth}
G.~Placidi, M.~Alecci, and A.~Sotgiu.
\newblock Theory of adaptive acquisition method for image reconstruction from
  projections and application to epr image.
\newblock {\em Journal of Magnetic Resonance}, pages 50--57, 1995.

\end{thebibliography}

\end{document}